# Challenges for an Ontology of Artificial Intelligence

*Scott H. Hawley*

*Department of Chemistry & Physics, Belmont University, Nashville TN USA*

*Abstract: Of primary importance in formulating a response to the increasing prevalence and power of artificial intelligence (AI) applications in society are questions of ontology. Questions such as: What "are" these systems?  How are they to be regarded?  How does an algorithm come to be regarded as an agent?  We discuss three factors which hinder discussion and obscure attempts to form a clear ontology of AI: (1) the various and evolving definitions of AI, (2) the tendency for pre-existing technologies to be assimilated and regarded as "normal," and (3) the tendency of human beings to anthropomorphize.  This list is not intended as exhaustive, nor is it seen to preclude entirely a clear ontology, however, these challenges are a necessary set of topics for consideration.  Each of these factors is seen to present a 'moving target' for discussion, which poses a challenge for both technical specialists and non-practitioners of AI systems development (e.g., philosophers and theologians) to speak meaningfully given that the corpus of AI structures and capabilities evolves at a rapid pace.  Finally, we present avenues for moving forward, including opportunities for collaborative synthesis for scholars in philosophy and science.*

## Introduction

Society is undergoing profound transformation due to the increasing effectiveness and reach of artificial intelligence (AI) applications. Predictions and warnings abound that the ascendancy of AI poses an "existential threat" to humanity,[1,2,3] and not simply in the form of "killer robots" or sentient AIs rendering humans obsolete. On the contrary, a number of more "mundane" threats and opportunities exist, as AI applications are revolutionizing widely-held conceptions of personhood and work. Although much prior work exists from antiquity through 2010, recent advances in machine learning (ML) often exceed prior conceptions of AIs' capabilities.

Significant work on "Theology and AI"[4,5,6] predates the sweeping changes afforded by the successes of ML systems and the scale on which they are deployed. Many key theologians and ethicists were responding to "classic AI." While time-honored reflections on concepts of automation, personhood and agency still apply, what's new is that the scope and reach of AI applications in society, their effectiveness, their ability to learn and synthesize, and the kinds of tasks they perform all vastly exceed what was widely thought possible or even conceivable even ten years ago. As a result, our conceptions of AI have continued to evolve, and there is renewed interest in establishing a clear understanding of these systems and their implications for society. Fundamental questions continue to be asked, such as: What is AI? How are such systems to be regarded? Is it appropriate to ascribe agency to algorithms?

Derek Schuurman argues for the primacy of ontology[7] as a precedent to addressing issues of application: "Once we have established the ontological question of who we are and what machines are, we can start asking the questions about the best way to move forward, including questions about the appropriate use of AI."[8] Regarding something on the basis of what it *is* is consistent with a traditional philosophical orientation that says things act in accordance to what they are, *i.e.*, their ontology. There is a sense of immediacy to this approach in our modern technological society, for as George Grant remarked, "technology is the ontology of the age,"[9] and the particular technology increasingly touted for its potentially transformational character is AI. Andrew Ng expresses this in the claim that "AI is the new electricity,"[10] in other words, that AI is poised to empower and revolutionize all areas of society.

Some may regard the matter of ontology to be irrelevant, that one needs only to adopt an instrumentalist viewpoint of studying the interactions between humans, machines and other *actants* in the form of an Actor-Network Theory,[11] or that the fundamental understanding of AI will result exclusively from rigorous development in the wider context of Human Computer Interaction (HCI). While these views have merit — indeed we may arrive at the need for treatment in terms of HCI — there may be significant value in investigating what AI is *per se*. Such an account will face a few challenges which we will describe. A starting point for these challenges can be seen in the following biographical observations.

**Prelude: The Joy of the Creator**

I confess that early in my studies in ML, I found myself blurring the lines regarding the ontology of the ML applications I would come across. Although I tended to be one who would be quick to point out the errors of others who anthropomorphize "intelligent" systems of various kinds, I found myself doting on, even cheering on the very "bots" that I had written from scratch, as I watched them grow in ability to perform some task.

I am a latecomer to the field of ML, although multidimensional nonlinear optimization problems were part of my training and my Ph.D. as a computational astrophysicist. The underlying techniques for the centuries-old problem of "curve-fitting" in the physical sciences amounts to a large class of ML problems. In fact, the problem of fitting a line to a set of data points is so fundamental that many ML curricula and tutorials use it as a foundational example for neural networks and/or evolutionary algorithms.[12,13,14,15]

Yet I never got *as* excited about watching my equation-solving numerical methods converge to a solution as I have watching simple ML toys "learn." The first example that hooked me as an "ML enthusiast" was a tutorial by Andrew Trask in which a Recurrent Neural Network (RNN) "learns" to do binary addition.[16] Seeing this "bot" start from nothing, making mistake after mistake but gradually improving, until finally achieving mastery, lit a fire of eagerness and curiosity and in me which continues. It is not obvious to me why this is the case: I had written numerous iterative-refinement solvers over the years (*e.g.*, using Newton's method), and yet for these I never made the cognitive jump to regard these systems as "learning;" I never anthropomorphized them.

Objectively, this RNN system is merely translating a series of binary inputs to a series of binary outputs by successively approximating some multidimensional mapping function, but it kindled in me a *joy*, a sense of having created something (even though the code was Trask's!), and that something was a tiny *agent*.

Where did this joy and attribution of agency come from? Was it born of *ignorance* about "what's really going on under the hood"? Only partly, for I also painstakingly recreated the code's matrix operations using a large Excel spreadsheet. As I did so, my enthusiasm diminished somewhat, but mostly because the process was indeed painstaking. Sharing the original code with students two years later, I still experienced excitement and a sense of wonder similar to my first encounter.

There seems to be a qualitative difference between regarding something as a mathematical operation and attributing intelligent agency to it. Part of this has to do with the ways we typically define intelligence.

## Challenge 1: Changing Definitions of AI

The term "artificial intelligence" has a long and varied history and tends to mean different things to different people. For some, it means nothing short of being able to perform any cognitive task that a human being can. For others, demonstration of very limited and task-oriented competence may suffice. Still, for others, AI is a marketing term chosen in recent years either intentionally or reluctantly, by those researchers who admit that "statistics" garners the least amount of enthusiasm or "buzz" from the general population, with "machine learning" generating greater buzz, leading up to "artificial intelligence" which may invite media frenzy. The various re-brandings of AI concepts with different terminology throughout its history may further obscure what sort of AI one is talking about. As UC Berkeley professor Michael Jordan notes, "The current public dialogue about these issues too often uses 'AI' as an intellectual wildcard, one that *makes it difficult to reason about the scope and consequences* of emerging technology"[17] (emphasis mine).

Although the concept of "machines that can think" has existed for many years, and initially was investigated in depth by Alan Turing, the term "artificial intelligence" was coined by John McCarthy, who organized the first artificial intelligence conference at Dartmouth College in 1956 for the purpose of organizing an effort to create human-like intelligence in a machine. McCarthy used this terminology to distinguish this line of research from the preexisting field of Norbert Wiener known as "cybernetics," which was defined in terms of control and communication of animal and machine systems. The field of ML arose primarily in a cybernetics context, rather than in trying to simulate human thought, but given that the application goals of many ML systems involve performing human-like tasks, the connection with AI is a close one. ML is now commonly regarded as a subset of AI, and so we will follow similar usage.

Writing a succinct definition of AI is a process with so many non-unique outcomes that there exist catalogues of various definitions,[18] even classified according to the principles underlying each definition.[19,20]

The source of variation in defining "artificial intelligence" lies more in the "intelligence" part than the "artificial" part.[21] Some experts take a minimal definition, defining intelligence as "doing the right thing at the right time,"[22] or any adaptive system (including evolution by natural selection[23]). Others cast it in terms of either thinking or action, with the goal of either mimicking humans or meeting some rational standard.[24] Components of intelligence may or may not include awareness, perception, reasoning, planning, and/or goal-setting. Consciousness, another concept with no clear consensus of meaning, is missing from many definitions of intelligence, and thus may be regarded as non-essential.

These variations are reflected in the choices of terminology which AI researchers have employed over the years to describe their work, often developing specialized nomenclature to distinguish their approaches from others. A few such specific terms are worth covering, as they will inform later discussion.

The terms "classic AI" and "Good Old Fashioned AI" (GOFAI)[25] refer to systems which employ human-programmed expertise and symbolic representations to behave in certain ways, using so-called "handcrafted knowledge."[26] This approach is exemplified in "expert systems" which often operate on the basis of hard-coded decision trees. Intuit's TurboTax program is a well-known example of this: by asking the user a series of questions, the algorithm is able to do the work of a tax accountant.[27] One important class of GOFAI consists of game agents such as the IBM chess system DeepBlue[28] which famously defeated Grandmaster Gary Kasparov in 1995.[29] Another relevant GOFAI example is Joseph Weizenbaum's computerized "therapist" ELIZA,[30] which used a series of pre-programmed patterns to mimic human dialogue.

In contrast to GOFAI, ML systems generally operate in numerical rather than symbolic ways, performing statistical inference from large datasets using an iterative optimization procedure that produces effects akin to human learning. Whereas GOFAI suffered from "brittleness" or catastrophic failure for small deviations outside the prescribed domains of their rules, the numerical nature of ML systems tends to allow for more "graceful degradation — in which imperfections in the data lead to proportionally imperfect but often acceptable performance."[31] To emphasize the statistical nature, some researchers prefer to use the term "statistical learning" for such systems.[32,33] The ML successor to DeepBlue was Giraffe[34] and later AlphaZero,[35] both of which achieved chess mastery purely by learning from self-play. The latter not only defeated human experts but demolished the highest-rated expert-system chess program.[36]

This ability for ML approaches to outperform classic AI systems has been seen dramatically in the results of trained neural network models which exceeded the performance of human-programmed algorithms in domains such as image and speech recognition,[37] and currently comprise the "best in class" solutions for many tasks.[38] This success has become so remarkable that the use of machines for tasks such as image recognition or speech synthesis are increasingly referred to by the tasks themselves, *e.g.*, "facial recognition" rather than "AI." We will discuss this de-assignment of the moniker "AI" further as part of Challenge 2: The New Normal, below.

The task-specific nature of applications of classic AI and ML to date have also caused disagreement over whether these constitute true AI.[39] Some choose to use the term "weak AI" or "narrow AI" for such applications, to distinguish them from "strong AI"[40,41] or "artificial general intelligence" (AGI) which involves mimicry of human-like performance across all cognitive domains. A vast amount of speculative fiction has been written about AGI, but so far, the speculation has vastly outstripped reality; we still have yet to see any computer code implementing a significant part of an AGI system. Even recent sensational claims to the contrary[42] seem upon closer inspection to fall short of the AGI ideal. One reason for this imbalance of fiction to reality will be discussed later in Challenge 3: Anthropomorphism.

As we described in the Prelude, ML algorithms have much in common with iterative approximation techniques which have been known since the days of Isaac Newton. Given the close association between ML and AI, often expressed mathematically as ML ⊂ AI, this means that longstanding data analysis techniques used throughout the sciences are becoming re-branded as ML and hence AI, often in order to take advantage of the current cycle of "AI Hype."[43,44] Algorithms such as fitting a curve to a set of data points were previously not regarded (by many) as constituting AI, and yet their implementation as the core methods of AI applications has brought such techniques to the forefront of discussions on AI — indeed, it has often been remarked that the "new wave" of highly successful ML systems applied essentially similar statistical techniques to those from years past, but with the benefit of vastly greater stores of training data (made possible by the internet).[45] Thus the underlying ontology of what the algorithm *is* may not be as important for determining the appropriateness of the label "AI" as the *intended use* of the system. Because of the ambiguities associated with the term AI, some researchers prefer to avoid its use and constrain their discussions to the specific ML algorithms involved —- Random Forests, Hidden Markov Models, Non-negative Matrix Factorization, Independent Component Analysis, Naïve Bayes, Gaussian Processes, (Artificial) Neural Networks, Deep Learning, etc. This specificity is useful from a technical perspective, but ontologically these algorithms are qualitatively of the same kind. What is ontologically relevant for all these is that since a deployed ML system is a function of (and thereby not easily separable from) its training dataset and even the particular starting point for the training procedure,[46] this means that an ML-based AI "is" not merely the algorithm and its intended use, but also the dataset used to train it.

In addition to the various uses and terminology just described, the AI definition which seems to be most applicable in regard to the implications of AI on the development of society is one which exists on the level of near folklore:

"AI is a computer doing what we *used to think* only a human could do."[47]

This "folklore" definition seems to capture the way that researchers and the public regard both new AI technology (*i.e.*, when a task or problem is being attempted or when a new paradigm is introduced) and old AI technology (*i.e.*, after some time has passed since a problem is largely regarded as "solved").  We will explore this in greater detail in the next section; it is the "used to" part of the "folklore definition" that leads us to Challenge 2 for an ontology of AI.

The central point of this section is not the mere observation that there exist a variety of possible definitions for AI.  While the fact that AI is not a monolithic, universal concept does pose some difficulty, the principal challenge arises from the fact that the *scope* of what is considered to "count" as AI is continually undergoing revision.  One might assume that this scope is monotonically increasing, however in the next section we note a mechanism by which this scope can also shrink, and thus the overall landscape of "what is [regarded as] AI" is in a state of flux.

## Challenge 2: The New Normal

The "folklore" definition of AI resonates with remarks by Douglas Adams on the "normalization" (sometimes referred to as "reification") of technologies:

> I've come up with a set of rules that describe our reactions to technologies:
> 1.  Anything that is in the world when you're born is normal and ordinary and is just a natural part of the way the world works.
> 2.  Anything that's invented between when you're fifteen and thirty-five is new and exciting and revolutionary and you can probably get a career in it.
> 3.  Anything invented after you're thirty-five is against the natural order of things...and the beginning of the end of civilisation as we know it until it's been around for about ten years when it gradually turns out to be alright really.[48]

Now that speech recognition systems are successfully employed in smartphones and smart speakers with high degrees of accuracy, many members of the public may not regard speech-to-text conversion (itself) as "AI," even though previously such systems were considered by many to constitute AI.  Even if such applications arose via training of sophisticated ML systems which themselves may count as AI to the researchers developing them, the normalization, ubiquity and reification of applications like Siri and Alexa have allowed members of the public to regard speech-to-text conversion as simply a tool or a task, without ascribing any intelligence to the system performing it.  One may inquire: Now that systems are able to learn from "experience," do people still regard Expert Systems as AI?  Or do they say, "That's just…" (*e.g.*, a set of nested if-then statements).  When one hears the phrase, "That's not really AI, that's just…" — an ontological assertion — it may indicate the speaker reserves "AI" for AGI, or it may indicate a change in attitude, *i.e.* a re-estimation of the worthiness of the "AI" label in favor of a more specific, mechanistic label which focuses on the task being completed, without regard for any intelligence used to complete the task.

This means that the term AI, even within the limited context of ML, is a "moving target."  The challenge this implies for developing an ontology of AI is that the usage of the term may be inseparable from whatever the current state of technology is when the term is being applied.

The first two challenges for a clear ontology of AI may be seen to involve the demarcation of AI in both conceptual and linguistic terms. One may rightfully raise the question of which community's conceptions and language are most relevant: the algorithm developers, the technologists who apply and deploy them, the journalists who break news about these developments, the general public who must come to terms with them, the philosophers who wish to make sense of them, or notably the theologians who wish to respond to them in the context of Biblical teaching?  Surely, one may argue, the general public and journalists often misquote or misapply the ideas of more rigorous thinkers on a variety of topics, and AI should be no exception. But the preceding observations are not limited in scope to any one particular subculture, and there is often an interplay of influence between these various groups, and the larger topic of "AI, Ethics and Society" merits discussion among all of them.  Even the most careful thinkers, it is argued, may have no way to avoid basic human tendencies that obscure attempts at clearly demarcating AI from other related concepts.  One such unavoidable tendency is that of anthropomorphism.

## Challenge 3: Anthropomorphism

The tendency to ascribe human faculties and/or intentions to entities in the world (animals, machines, objects, "forces of nature") has existed since antiquity.  Francis Bacon observed that it often impedes our understanding of the natural world, as what he called "The Idol of the Tribe": "For it is a false assertion that the sense of man is the measure of things."[49] Put differently, anthropomorphism amounts to a "cognitive bias"[50] and as such impedes one's ability to regard things as they are — *i.e.*, ontologically. Despite its association with unenlightened eras, anthropomorphism occurs even today – perhaps even more so than previously. As Waytz *et al.* observed, "Although [anthropomorphism is] commonly considered to be a relatively universal phenomenon with only limited importance in modern industrialized societies—more cute than critical—our research suggests precisely the opposite."[51]

Anthropomorphism appears as the "go-to" model or metaphor by which humans initially seek to understand new phenomena — the "hammer" we try to apply to many "nails," if you will. Beth Singler of the Faraday Institute has said anthropomorphism arises "because we are social beings who need to place the things around ourselves into a social scheme that makes sense of them."[52] It is widely speculated that our cognition is biologically optimized to process our "local world" which is predominantly a social one. A common sentiment is that it is "hypothesized to have evolved because it favored cooperation among early humans."[53] So strong is the tendency to project human-like qualities onto other things, that it is regarded as unavoidable. As mechatronics researcher Dr. Fumiya Iida has described, "Anthropomorphization is [an] incurable disease for human[s]."[54] Anthropomorphism appears to be more likely to arise in situations for which detailed operational knowledge is not available, or when novel unexpected emergent behavior arises, such as in the case of certain moves by AlphaGo.[55,56,57]

Anthropomorphism plays a key role in the *design* of AI systems, and even in the conception of AI. The earliest formulations of the concept of AI are anthropomorphic. The "Turing Test"[51] is built around the model of human intelligence: can a machine communicate in such a way as to fool a human into regarding it (the machine) as human? McCarthy's goal of the Dartmouth conference[58] was explicitly human-centric. Beyond that, anthropomorphism is found to serve a utilitarian purpose in design, which "can be used today to facilitate social interactions between humans and a new type of cooperative and interactive agents – social robots."[59] This means that it can allow for more intuitive use of such robots, particularly in "caregiving" applications such as intervening in the development autistic children,[60,61] and some care of the elderly.[62,63] It has also been warned that the anthropomorphic urge could be hijacked to create inappropriate bonding with artifacts, and thus ethical design should provide transparency to avoid such misuse.[64]  Concerns about the inappropriate use of anthropomorphic aspects of AI led Weizenbaum, creator of the ELIZA "psychotherapist," to later oppose the use of such systems in "interpersonal" settings:

> "I would put all projects that propose to *substitute* a computer system for a human function that involves *interpersonal respect, understanding, and love* in the same category. I therefore reject [Kenneth] Colby's proposal that computers be installed as psychotherapists, not on the grounds that such a project might be technically infeasible, but on the grounds that it is immoral."[65](emphasis mine)

The effects of the cognitive bias of anthropomorphism are manifold.  Robert Wortham observes that it can result in "moral confusion about the status of robots in particular, and artificial intelligence more generally."[66] This confusion can involve questions of whether robots should have rights,[67] whether AI systems should be granted status as legal persons,[68] and in general whether humans have a responsibility toward robots, so-called "moral patiency."[69] Moral patiency of machines is regarded as such a serious danger that Joanna Bryson states forcefully, "We are therefore obliged not to build AI we are obliged to."[70] It is anthropomorphism which is identified as a key obscuring factor contributing to misperceptions of moral agency and/or patiency, as Wortham continues: "There are serious concerns that our anthropomorphism and misunderstanding of the nature of robots extends so far as to attribute them either moral patiency, moral agency, or both."[71,72,73]  A further common effect is that of "overidentification,"[74] in which humans may ascribe additional human attributes to machines, based on performance at tasks of logic and language. That is to say, having observed a system performing tasks of logic and language, there is a common human tendency to ascribe or *project* a host of additional cognitive and behavioral faculties onto the machine. This *extrapolation* by the user may be unwarranted, such as in the example of a "self-driving" car which can stay in its lane well and thus come to be regarded as an excellent driver, but can be thwarted by the appearance of a bicyclist[75] or a lane division[76] and lead to death; overidentification is a likely contributor to driver inattention in such cases.   Finally, anthropomorphism has the effect of making it all too easy to write (yet more) fiction about AGI.  This can distract conversations from real, immediate dangers and opportunities, to speculations on severely underdetermined scenarios set in the far future. As Andrew Ng recently lamented, "AI+ethics is important, but has been partly hijacked by the AGI (artificial general intelligence) hype. Let's cut out the AGI nonsense and spend more time on the urgent problems: Job loss/stagnant wages, undermining democracy, discrimination/bias, wealth inequality."[77]

Existing in a "dual" relationship to anthropomorphism is the tendency to dehumanize (or objectify), an *ontological error* whereby the personhood, individuality and value of human beings are denied and replaced with a regard for humans only as things. In committing this error, we move from the "I-Thou" mode of relation identified by Martin Buber,[78] to one of "I-It." In an AI context, dehumanization arises in a variety of ways. Firstly, it may be explicitly stated, in a naturalistic approach to the so-called "mind-body problem," that humans are merely machines and that the mind is not simply like a computer, it is a computer.[79] In contrast, in the words of Schuurman, "A Christian perspective accounts for reality as extending beyond the physical world to include a spiritual realm. This ontological starting point will reject the reductionistic notion that humans are simply complex biochemical machines, while still affirming the value of the physical world."[80] Secondly, dehumanization can arise as a result of the anthropomorphism of artificially intelligent systems. Bryson states it thusly: "In humanising [robots], we...further dehumanise real people."[81] The other primary avenue for dehumanization arises from its utility in modeling human behavior, for applications such as recommendation systems and targeted marketing, and for *manipulating* human behavior. This was evidenced in the news of Facebook's deliberate attempts to make their application more addictive, referring to people as "eyeballs."[82] The ontological error of viewing humans as machines can have a series of ethical consequences, such as in the area of employment: The extent to which we view humans mechanistically suggests the extent to which we will automate people out of jobs. Christians have historically opposed the tendency to dehumanize and objectify, on the basis of love of one's neighbor and the doctrine of *imago dei*. This is an area in which Christians can continue to have a significant witness to the larger society, as the temptations to dehumanize are likely to increase along with the scale of deployment and success of AI systems at performing various tasks. This also presents opportunities for partnership with secular individuals and institutions dedicated to the ethical use of AI, as Christian positions are often in agreement with secular ones, such as in opposing the dehumanizing implications of sex robots,[83] or of AI-empowered surveillance technology and the utility of classification systems for enabling oppressive government practices.[84]

**Avenues for Moving Forward**

While a rigorous ontology of AI may be difficult, it has not been shown to be impossible. On the other hand, it may not be necessary, as alternative approaches are available. Given that questions regarding AI are invariably bound up with questions of humanity, it may be that AI is not a distinct concept that can be well-demarcated from humanity, and thus a broader context of Human-Computer Interaction (HCI) may be a more fruitful avenue. Alternatively, an "instrumentalist" approach such as Actor-Network Theory,[85] that would focus only on what AI *does* in its interactions with other parts of a larger system, may provide a more efficient route to answering questions of application and appropriate use. Finally, a "process philosophy" which regards things not as they *are* but how they undergo *change* — which amounts to an ontological position albeit with a different emphasis from the traditional one — may prove profitable.

It is not the intent of this paper to evaluate the relative merits of these approaches in comparison to an ontological approach, however one key question in such an evaluation would be: Does an ontology of AI "get you anything" that these other approaches do not? An answer to this may lie in current discussions of explainability and transparency. A "black box" system which is known only via its exterior interactions is unlikely to garner public trust[86] and likely fails to meet the "right to explanation" requirement laid out in the European Union's General Data Protection Regulation (GDPR).[87] There are methods for probing the internal logic of black boxes with the goal of explainability,[88] however these are not applicable in all situations. In general, the issue of transparency is not a simple one, for naïvely manifesting the totality of what an algorithm *is* — by exposing its source code and, for example, the internal weights of a neural network, and also its (potentially biased) training dataset[89] — does not constitute an explanation, and designers of systems with transparency in mind must consider the *level* of detail shared so as not to overwhelm the user, and to provide transparency with the goal of understanding in mind.[90] Such a level of detail would need to be chosen according to the intended users, and amounts to a kind of user interface design. In this case, transparency may be regarded as being more consistent with instrumentalism than ontology, because the emphasis is on clarifying what the system is *doing* rather than what it is. However, the function of transparency is to foster "the ability of a naive observer to form an accurate model of a robot's capabilities, intentions and purpose,"[91] for goals which include clearly demarcating the *ontological* difference between a machine intelligence and human, and in so doing, to mitigate the effects of anthropomorphism discussed earlier.

Another notable avenue, which is both a form of ontology and an alternative orientation, is the functionalist approach to AI-ethics employed by Joanna Bryson.[92,93] In this case there is an ontology that posits the inequality of the human and the machine, however with the goal of preservation of the social order rather than the affirmation of any metaphysical significance to the individual (*i.e.*, without an *imago dei*). It is notable that, once again, although the functionalist approach rests on a different foundation than Christian ethics, many of the implications of the former for "AI, Ethics and Society" are sufficiently in alignment with traditional Christian moral and ethical positions that significant opportunities for partnership exist between Christians and those operating from such "secular" standpoints.

Finally, we note that the pace of advancements in AI, particularly in the area of ML, has become so rapid that dramatic announcements arise with a frequency of every few months, with six months being a common timescale for significant achievements. It is typical for successful methods to be superseded within a year or two, and the understanding of their implications to require some revision. Thus the peer-review process in the ML field occurs more in the form of conferences than in journals (whose longer review time can impede dissemination). In such a swiftly changing landscape, it is possible for those without immediate connection to the technical field to make statements which no longer apply, for example "AI doesn't do X, it merely does Y," only to be corrected that indeed "AI now does X, as of six months ago." So for Christian philosophers and theologians, it is recommended that they form partnerships with those in the technical sectors of academia and/or industry in order to stay current — and therefore relevant. This assumes that there will exist participants in the technical domains who are interested in partnering for the purpose of Christian scholarship, and thus we see, as with other areas of science, the need for Christians to enter such fields and perform excellent work with diligence and integrity.

**Conclusions**

In response to interest in establishing an ontology of AI, we have not achieved this goal, but have raised awareness of three challenges which can hinder dialogue and obscure clear thinking about what AI is.  These challenges are significant and make the establishment of a clear ontology of AI more difficult.  The first challenge is the various and changing ways in which AI is defined in terms of the research community and society at large, and the evolving scope of what sorts of algorithms "count" as AI.  The second challenge is a result of the widespread successful deployment of some AI systems, when they reach the point where the previously-challenging tasks they perform come to be reified and regarded as simply "normal," such that the frontier of what is regarded as AI by the general populace advances toward more difficult problems.  These first two challenges are more than mere semantic objections: in the words of theologian Michael Burdett, "the language we use is important because it manifests our ontological commitments."[94] Conversely, as Mary Midgely has argued, our language not only exposes but tends to shape our ontological commitments.[95]  The final challenge is that of the unavoidable human tendency to anthropomorphize, which yields a cognitive bias that can manifest in ways such as projecting moral agency and/or patiency toward machine intelligences. The advancing performance of AI at tasks of language and logic means that the overidentification of human attributes with AI is likely to evolve as well.  This is a challenge but also the reason why developing a clearer ontology for AI is an important undertaking.  In order for scholars in theology and philosophy to keep pace with the rapid changes in the technical performance, conceptions and scope of AI systems, it is recommended that collaborative partnerships be formed with active technical practitioners of AI systems development.

**Acknowledgements**

The author wishes to thank the following persons for helpful discussions: Michael Burdett, Alister McGrath, Tommy Kessler, Andy Watts, William Hooper, Beth Singler, Andreas Theodorou, and Robert Wortham.  Sponsored by a grant given by Bridging the Two Cultures of Science and the Humanities II, a project run by Scholarship and Christianity in Oxford (SCIO), the UK subsidiary of the Council for Christian Colleges and Universities, with funding by Templeton Religion Trust and The Blankemeyer Foundation.